%% file: main.tex
\documentclass[apj,twocolumn,twocolappendix,numberedappendix]{openjournal}

\usepackage{newtxtext,newtxmath,enumitem,multirow}
\usepackage[utf8]{inputenc}
\usepackage[T1]{fontenc}
\usepackage[breaklinks,colorlinks,allcolors=blue,citecolor=blue,urlcolor=blue]{hyperref}
\usepackage{orcidlink}
\usepackage{ctable}
\usepackage{array}

\usepackage{makecell}

\renewcommand{\arraystretch}{2}

\pdfoutput=1

\shorttitle{Transition metals and the search of Life}
\shortauthors{Covone \& Giovannelli}

\begin{document}
\title{
%Stellar metallicity is a key parameter for the search of Life in the Universe \\
Transition metal abundance as a key parameter for the search of Life in the Universe 
\vspace{-1.5cm}}

\author{Giovanni Covone\,\orcidlink{0000-0002-2553-096X}$^{1,2,3,\star}$}
\author{Donato Giovannelli\,\orcidlink{0000-0001-7182-8233}$^{4,5,6,7}$
\vspace{1mm}}

\affiliation{$^1$ Department of Physics “E. Pancini”, University of Naples Federico II, Naples, Italy}
\affiliation{$^2$ INAF -- Osservatorio Astronomico di Capodimonte, Salita Moiariello 16, I-80131, Napoli, Italy}
\affiliation{$^3$ INFN, Sezione di Napoli, Via Cintia, I-80126 Napoli, Italy}
\affiliation{$^4$ Department of Biology, University of Naples Federico II, Naples, Italy }
\affiliation{$^5$ National Research Council – Institute of Marine Biological Resources and Biotechnologies - CNR-IRBIM, Ancona, Italy} 
\affiliation{$^6$ Marine Chemistry\& Geochemistry Department - Woods Hole Oceanographic Institution, MA, USA}
\affiliation{$^7$ Earth-Life Science Institute, Tokyo Institute of Technology, Tokyo, Japan}
\thanks{$^\star$\href{mailto:kravtsov@uchicago.edu}{giovanni.covone@unina.it}}

% These dates will be filled out by the publisher
%\date{Accepted XXX. Received YYY; in original form ZZZ}

\begin{abstract}
The search for Life in the Universe generally assumes three basic life's needs: I) building block elements (i.e., CHNOPS), II) a solvent to life's reactions (generally, liquid water) and III) a thermodynamic disequilibrium. It is assumed that similar requirements might be universal in the Cosmos. On our planet, life is able to harvest energy from a wide array of thermodynamic disequilibria, generally in the form of redox disequilibrium. 
The amount of different redox couples used by living systems has been estimated to be in the range of several thousands of reactions. Each of these energy yielding reactions requires specialised proteins called oxidoreductases, that have one or more metal cofactors acting as catalytic centres to exchange electrons. 
These metals are de facto the key component of the engines that life uses to tap into the thermodynamic disequilibria needed to fuel metabolism. The availability of these transition metals is not uniform in the Universe, and it is a function of the distribution (in time and space) of complex dynamics. 
Despite this, Life's need for specific metals to access thermodynamic disequilibria has been so far completely overlooked in identifying astrobiological targets. 
We argue that the availability of at least some transition elements appears to be an essential feature of habitability, and should be considered a primary requisite in selecting exoplanetary targets in the search for life. 
\vspace{2mm}
\end{abstract}

\keywords{astrobiology, planets and satellites: atmospheres, planets and satellites: general}

\maketitle

\section{Introduction}

The notion of habitability plays a key role in the search for Life in the Universe (see Appendix A for definitions of some of the terms used in this paper) % perspective). 
In astrobiology, habitability is usually defined as “the ability of an environment to support the activity of at least one known organism” \citep{Cockell_2016}. 
The notion of habitability is given by the set of known physical and chemical conditions necessary for the emergence and the evolution of life as we know it. However, the knowledge of these conditions is still approximate. 
We have a primitive knowledge of Life as a natural phenomenon in the Universe, and the true limits of life on Earth are still largely undefined \citep{Merino_2019}. Hence, beside a universal definition of habitability \citep{Cockell_2021}, we need a conceptual tool which might guide the selection of the most promising targets \citep{Tasker_2017}. 
Pre-selection of targets is unavoidable, as astronomical searches for biosignatures are very expensive in terms of telescope time and data analysis effort. 
As both resources are strongly limited, an effective strategy in searching for life needs a reliable selection of targets, among the ever increasing number of known exoplanets.  

The current notion of habitability is based on our present-day knowledge of the basic needs for the existence of life on Earth. In particular, the search for life in the Universe generally assumes the three following requirements: a) building block elements, i.e. molecules composed of the so-called CHNOPS elements (Carbon, Hydrogen, Nitrogen, Oxygen, Phosphorus and Sulphur, the major macromolecules building elements used by life); 
b) a source of free energy which can sustain thermodynamic disequilibria \citep{Hoehler_2007} and c) liquid water, acting as a universal solvent to life’s reactions \citep{Westall_2018}. 
There is no lack of the first two ingredients in the cosmos \citep{Randich_2021}, and recent data suggest that water is also abundant \citep{Hanslmeier_2011}. 
Here, we argue that the availability of transition metals (here the term "metal" is always used in the traditional non-astronomer sense) is a strong constraint on the distribution of Life in the Universe and should be systematically considered while selecting the most promising candidates for the search for extraterrestrial life.

Current data suggests that CHNOPS elements are present in a large number of organic chemical species in the interstellar medium, in the star-forming regions and in protoplanetary disks \citep{Nomura_2021}, making the basic building blocks of life a non limiting factor in the emergence 
%and evolution 
of Life in the cosmos. 
The energy needed for biomass synthesis is also abundant in the Universe. Stars are a long-lasting and abundant source of free energy sustaining dynamic planetary atmospheres and the eventual biospheres. For instance, free energy flux is not a limiting factor for production of biomass by means of (hypothetical) oxygenic photosynthesis on exoplanets around main-sequence stars hotter than about 3000 K \cite[see for instance][]{Covone_2021}, and geological processes might sustain life through geochemical and thermal disequilibria without the need for light \citep{Merino_2019}. Hence, the first two necessary ingredients for life appear to be abundant on the cosmic scene and potentially in all planetary systems. Also liquid water is likely not a limiting factor. 
Recent observational data and theoretical studies suggest that the presence of liquid water (including subsurface oceans), might be more abundant in the cosmos than previously estimated \cite[e.g.,][]{Noack_2017}.

Considering all of the above, astronomers define the “habitable zone” as the circumstellar region where a rocky planet with an Earth-like atmosphere could have liquid water on its surface \citep[e.g.,][]{Cockell_2016}. 
It is also currently accepted that the definition of habitability can be extended by specific planetary conditions that might support the presence of liquid water and temperatures conducive for life existence. 
For example, the presence of subsurface oceans on icy moons
(as Enceladus, Europa and others in the Solar System) makes 
them promising targets \citep{Kanik_2021}. 
Also the rocky subsurface of planets could be considered habitable given the vast subsurface ecosystems present on Earth \citep{McMahon_2013}. 

This set of conditions includes a large number of possible astronomical targets. For instance, recent work predicts that the forthcoming PLATO mission \citep{Rauer_Plato_2025} will allow to discover $24 \pm 4$ Earth-like exoplanets (i.e., planets with radius between 0.8 – 1.25 Earth radii) around FGK stars on long period orbits (between 250 and 500 days), roughly including the circumstellar habitable zone \citep{Matuszewski_2023}. 
We remark that FGK stars offer much better conditions for Life in terms of stability (i.e., frequency of high-energy flares) and X-ray and UV irradiance with respect to the most numerous M star dwarfs. If we also consider larger exoplanets (i.e., sub-Neptunes), this figure increases enormously: 
$172 \pm  15$ exoplanets with radii between 1.2 and 2 Earth radii %(so called super-Earths)
\citep{Matuszewski_2023}. 
This set includes Hycean worlds, that is, worlds hosting water oceans at their surface beneath an H$_2$-rich atmosphere. Hycean worlds broaden the generally considered range of habitability, in terms of planetary mass, radius and orbital distance (see for instance \cite{Rigby_2024}). 
The larger sizes of sub-Neptunes makes them favourable targets for atmospheric characterization and biosignatures search via transit spectroscopy. About one third of these systems could be around stars bright enough to allow follow-up for mass measurements (via Doppler spectroscopy) and atmospheric characterization (with the JWST), with larger exoplanets being favoured in this metric. 
Thus in a few years, discoveries from PLATO and the on-going TESS missions will likely exceed our current observational follow-up capacity. 
Prioritising promising targets using present-day knowledge of Earth's life appears necessary for an effective astronomical search for biosignatures.

\section{The role of stellar metallicity in habitability}

Considerations on the availability of heavy elements in the context of habitability have focused so far almost exclusively on their role in providing the “bricks” for the formation and composition of planetary systems. Elements heavier than hydrogen and helium are essential in this process: indeed, the very first generation of stars probably could not produce planets as we observe today, as most of the heavier elements have been formed through nucleosynthesis processes connected with star evolution \citep{Johnson_2019}. 

Star metallicity $Z$ (i.e., the abundance by mass of chemical elements heavier than helium relative to the total mass of the star) appears to be correlated to the presence of giant planets \citep{Brewer_2018}, 
probably because the prestellar enrichment of heavier elements in the interstellar matter is necessary for the formation of the giant planets’ rocky cores (see \cite{Fischer_2005} and references therein). Statistical analyses of known exoplanets show no trend between rocky planets occurrence and host-star metallicity \citep{Zhu_2021}, albeit this might change in the future as more exoplanets are discovered. Indeed, a too low metal content in the protoplanetary disk is expected to inhibit the fast accretion of rocky bodies massive enough to start aggregating the gas in the circumstellar disk before this disappears in a few million years. The lower metallicity threshold necessary for rocky planet formation is still unknown. Theoretical models suggest that the first Earth-like planets might have formed from protoplanetary disks with metallicities $Z \gtrsim 0.1  Z_{\odot} $ \citep{Johnson_2012}.

Given the importance of metals in the formation of planets, it is not surprising that the relationship between star metallicity and habitability has been investigated in the past, focusing on the role of heavy metals on important characteristics of rocky planets. For instance, several authors have considered the role of heavy elements for planet formation and long-lasting geological activity \citep{Horner_2010}, the duration of the habitable (temperate) zone \citep{Danchi_2013} and the frequency of rocky planets in the habitable zone as a function of the metallicity \citep{Mulders_2016}. 
However, it appears that the connection between availability of heavy metals, and specifically transition elements, and Life has been so far neglected in the astronomical literature.

The ever growing list of exoplanets host-stars measurements hints at a widely varying metal-enrichments at a given stellar mass \citep{Morley_2017, Wakeford_2017}, and at a diverse C/Fe ratio between stars with low-mass planets and stars without planets \citep{Delgado_Mena_2021}. 
This suggests a possible relationship between CHNOPS distribution and metal availability that has yet not been explored in detail. Beyond  the importance of stellar metallicity in controlling planet formation, the availability of selected trans-iron transition elements might be fundamental in enabling the emergence and evolution of Life.

\section{Metal availability as a key control on Life emergence and evolution}

All life as we know it uses thermodynamic disequilibria to access energy used to drive biosynthetic chemical reactions. The “simple” act of converting an inorganic carbon molecule (most often in the form of CO$_2$) into a reduced organic carbon to build biomass 
(like the one performed during autotrophy to make 
glucose) requires energy to drive an otherwise thermodynamically unfavourable reaction. 
The necessary energy is usually harnessed from available thermodynamic disequilibria in the environment, coupling electron donors and acceptors in biologically-mediated redox reactions. 
For instance, hydrogen and carbon dioxide, commonly used as energy and carbon sources by methanogens in the reaction 

$$ 4 \, {\rm H}_2 + {\rm CO}_2 \rightarrow {\rm CH}_4 + \, 2 \, {\rm H}_2 {\rm O} $$
are a common and probably old redox couple used by life \citep{Martin_2008}. 
Sometimes complex energy sources are used to drive redox reactions, such as in oxygenic photosynthesis, in which photons are used to extract electrons from water \citep{Fischer_2016}. 
Life is thus capable of harvesting the necessary energy from the environment using complex molecular machineries. 
These nanoengines (Fig. \ref{fig:2}) are proteins (specifically, catalytic proteins known as enzymes, Appendix A) capable of catalysing redox reactions called oxidoreductases.

\begin{figure*}
    \centering
    \includegraphics[width=1\linewidth]{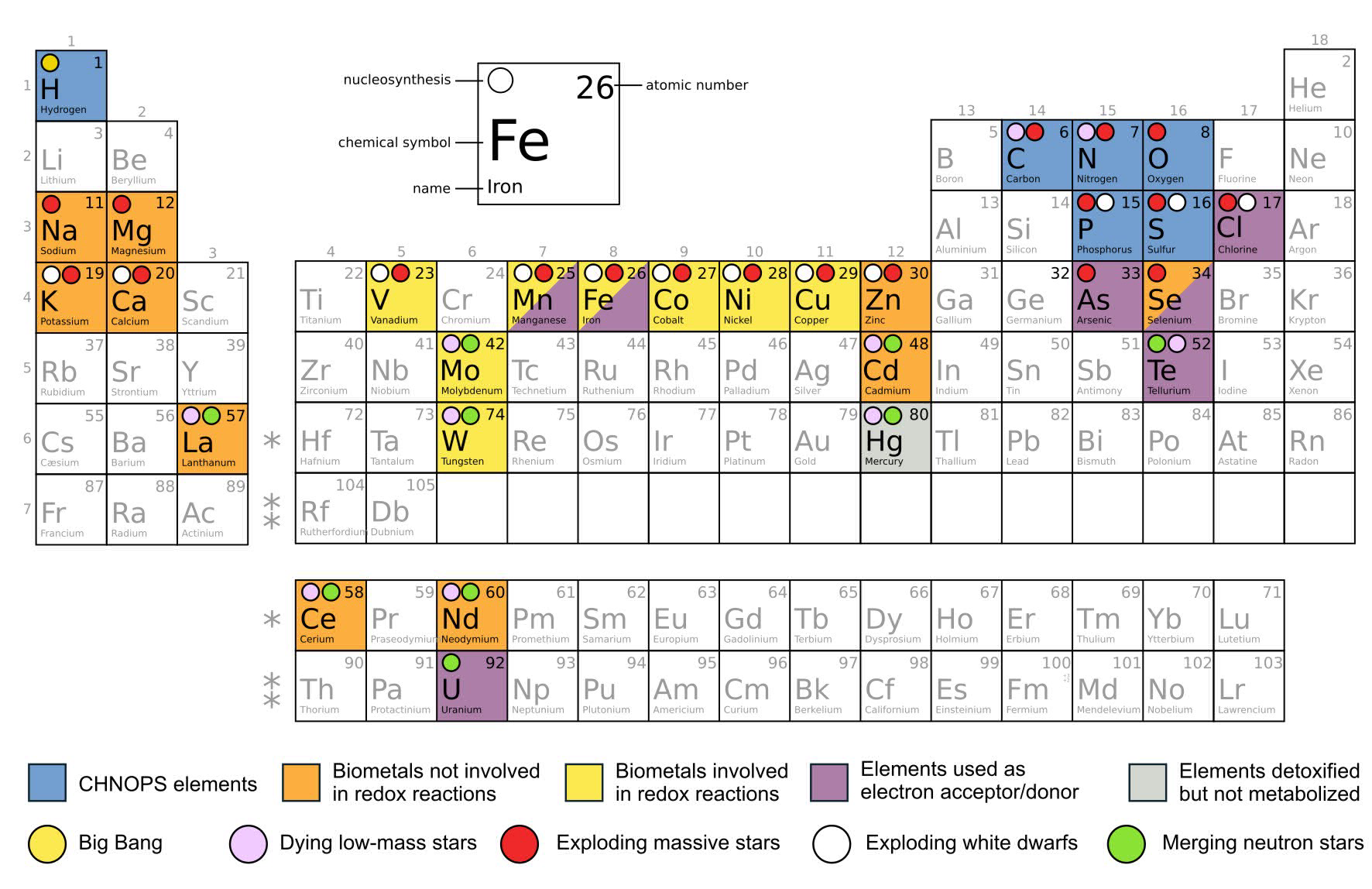}
    \caption{Periodic table of elements and their nucleosynthesis. Biological utilisation of the elements and biometals (in orange and yellow) and their nucleosynthesis processes in the Universe are indicated. 
    CHNOPS elements and elements used as electron donors/acceptors by biology are also reported. 
    Iron and Manganese are the only biometals that not only serve as cofactors, but might also be used as metabolic substrates for redox reactions.}
    \label{fig:1}
\end{figure*}

\begin{figure*}
    \centering
    \includegraphics[width=1\linewidth]{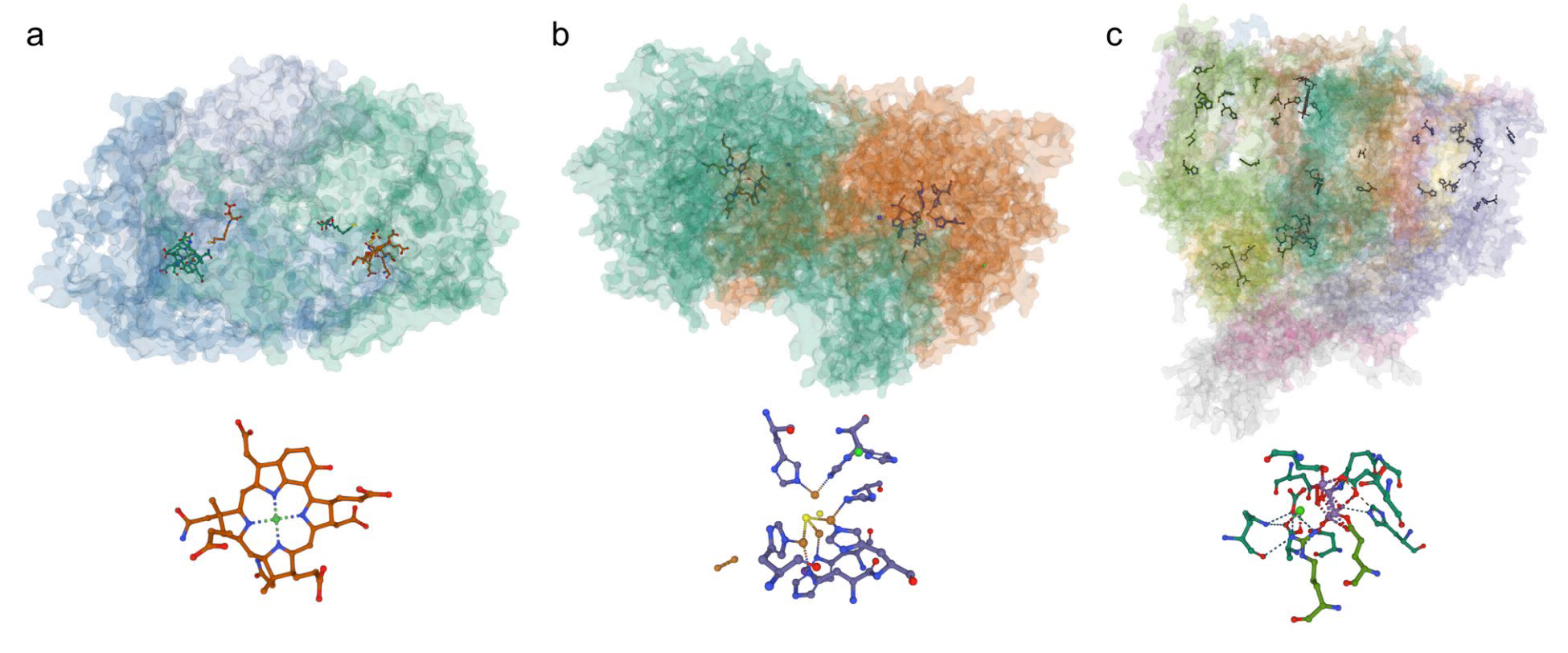}
    \caption{Example of metal containing enzymes involved in the production or utilisation of possible biosignature gases such as CH$_4$, N$_2$O and O$_2$ . The three dimensional structure of the protein is showed as a continuous molecular surface (with significant transparency applied for visualisation purpose) coloured according to the different amino acid chains, while the cofactors are shown using a ball and stick model within the enzyme and on their own coloured according to the type of atoms. (a), The key enzyme for methanogenesis, the Methyl-coenzyme M reductase from the archaea Methanopyrus kandleri (McrA, Protein Data Bank accession number 1E6V. Structures can be visualised and inspected at https://www.rcsb.org/). The nickel containing F430 cofactors are visible within the enzyme in two copies, together with other small ligands. The structure of the nickel (Ni2+) containing F430-cofactor (the nickel atom is coloured in green) responsible for the last reduction step required for methane production is visible below the enzyme. (b), The enzyme responsible for the production of nitrous oxide, nitrous oxide reductase from the bacterium Pseudomonas stutzeri (NosZ, PDB 3SBR) with the two copper containing cofactor visible within the enzyme. The structure of the copper sulphide cofactor (Cu$_4$S$_2$) and the dinuclear copper is visible below the enzyme. Copper atoms are coloured in orange while sulphur atoms are in yellow. (c), The complex multimeric structure of the Photosystem II from the cyanobacterium Thermosynechococcus elongatus (PSII, PDB 1S5L), where the oxygen evolving complex responsible for the splitting of water in oxygenic photosynthesis is located. All the chlorophyll and pigments have been removed for visualisation purposes. The structure of the oxygen evolving complex CaMn$_4$O$_4$ cofactor, the site of water splitting and oxygen generation is visible below the enzyme. Calcium is coloured in green, while manganese and oxygen are coloured in purple and red respectively. }
    \label{fig:2}
\end{figure*}

These enzymes are evolutionarily tuned to be able to transfer electrons to and from target molecules in a controlled way \citep{Falkowski_2008}, allowing life to take advantage of the energy associated with the reaction and to perform chemical (and often physical) work (Appendix B). Since redox reactions require the flow of electrons, oxidoreductases reaction centres need to be finely tuned in their midpoint electron potential to accept and donate electrons without dissipating unnecessary energy. The vast majority of known oxidoreductases uses a diverse set of metal containing cofactors to achieve this \citep{Giovannelli_2023}.
For example, oxidoreductases utilised to extract high energy electrons from hydrogen (a common electron donor in microbial redox reactions) contain generally a nickel-iron metal cofactor (see Fig. \ref{fig:2}), while many enzymes acting on oxygen (a strong electron acceptor respired by a large number of organisms) use instead copper containing cofactors to be able to gently pass electron to the strongly electronegative O$_2$ molecule. 
While most known enzymes use one or more first or second row transition metals (see Fig. \ref{fig:1}), the list of biologically important metal cofactors is growing \citep{Giovannelli_2023}. 
Only recently we have identified enzymes that catalyse the first step of methane oxidation (an important biosignature and greenhouse gas) using the rare earth elements lanthanum and cerium. The exact composition of the metal cofactors used by life is still unknown, however the majority of known oxidoreductases involved in key energy conserving reactions contain transition metals that include Fe, Cu, Mo, V, W, Co, Ni, Mg, Mn among others, see Fig. \ref{fig:1} \citep{Hay_Mele_2023}. 
These biometals  \cite[\textit{sensu}][]{Giovannelli_2023} are \textit{de facto} the key component of the engines that life uses to tap into the thermodynamic disequilibria needed to fuel metabolism. 

Life on Earth is entirely dependent on the environmental availability of biometals in order to sustain its growth. For example, primary productivity in the oceans can be severely limited by the availability of iron \citep{Wade_2021}, and similar controls might be exerted by diverse metals in other ecosystems \citep{Giovannelli_2023}. 
Metal availability might also have played a key role in influencing the evolution of Life
\citep{Wade_2021, Moore_2017}, as the availability of biometals has significantly changed over time as a results of key planetary transitions, such as the onset of plate tectonics
\citep{Edmonds_2018, Barnes_2021}, 
changes in volcanism \citep{Edmonds_2018, Liu_2021}
and the great oxidation event \citep{Anbar_2008}. 
Given the tight  link between the necessity to extract energy from thermodynamic disequilibria and the ubiquitous use of metals as key cofactor in accomplishing this, the role of metals in Life might be more fundamental than previously recognized \citep{Kacar_2020}, and should be considered as a key requirement to harness energy provided by the star or the geology of the planet. 

\section{Stellar metallicity as a key parameter in the search for Life}

While we search our solar system and beyond for targets of astrobiological interest, 
we argue that the availability of transition metals necessary to access a diverse array of thermodynamic disequilibria is an essential feature of habitability. 
% \textbf{Microlensing surveys can detect planets up to very large distances, highlighting the potential reach of exoplanet searches into diverse stellar populations with varying metallicities.}
%
Chemical elements heavier than helium (including all biometals) have been produced in detectable amounts relatively late in cosmic history \citep{Johnson_2020}, see Fig. \ref{fig:1}. 
Hydrogen and helium have been produced in the early primordial nucleosynthesis, while synthesis of heavier chemical elements could be possible in significant quantities only during several stages of stellar evolution, with the first generation of stars (the so-called Population III) appeared on the cosmic scene when the Universe was not younger than about 100-200 million years old \citep{Omukai_1998}. While details are still highly debated, this scenario is strongly supported by the measured chemical abundances in the oldest stars in the Galaxy and the molecular clouds \citep{Nomoto_2013}.

All biometals lighter than iron are the outcomes of fusion nuclear reactions in the core of massive stars (i.e., with a mass larger than about two solar masses) see e.g. \cite{Johnson_2019}. As iron is the most tightly bound nucleus (with binding energy 8.8 MeV per nucleon), it marks a separation between lighter elements that can be synthesised by the fusion of nuclei and heavier elements that are produced via nuclear fission processes. 
For instance, most of the heavy elements up to bismuth are produced in this way in Type-Ia supernovae explosions \citep{Nomoto_2013}. 
Supernovae explosions have also a key role in diffusing the nuclear ashes in the interstellar medium enriching molecular clouds.

As a consequence of the enrichment of the interstellar medium over time and the efficient mixing, at any given moment the distribution of the metal abundance in the Galaxy is very regular, with heavy-element abundances systematically decreasing outward from the centre. However, some observations show evidence for the survival of chemical inhomogeneities on smaller scales of about tens of kiloparsec. 
In particular, evidence comes from the large scatter in the star age - metallicity variations and from UV spectra along several directions. These small scale variations are likely due to the episodic infall of high-velocity, low-metallicity gas from the halo, as seen in small scale numerical simulations \citep{de_Avillez_2002}. 
For instance, variations of about one order of magnitude of the measured abundances have been found along the line-of-sights to a 
well-chosen sample of 25 bright stars \citep{De_Cia_2021}. Therefore, metal abundance is not expected to be very homogeneous on small scales and large variations could be found among the nearby stars.

Measurements of stellar chemical abundances are increasingly available for elements heavier than iron, allowing for the investigation of metal distribution in the host stars of distant planetary systems. 
In particular, the Apache Point Observatory Galactic Evolution Experiment (APOGEE, \cite{Weinberg_2019}) has completed an homogeneous, high-resolution and high signal-to-noise spectroscopic survey of the stellar populations of the Milky Way, allowing to obtain a detailed chemical composition for about 146,000 stars (see Figure 3). While the bulk planetary composition is controlled by the metal composition of the parent star \citep{Adibekyan_2021}, the crustal abundance of metals on a planetary scale is controlled by a large number of local factors. Processes like core formation, crustal differentiation, active geology and redox conditions of the planetary surface all influence the distribution and availability of CHNOPS elements and metals \citep{Wade_2021}, controlling their ultimate availability for Life. 

More in details, \cite{Wang_2019Icarus} found that elements with low condensation temperatures (lower than Earth's differentiation temperature) are depleted in the terrestrial environment compared to solar abundances. In contrast, elements with higher condensation temperatures match solar abundances.

Despite this, first order estimates of bulk planetary composition might be enough to determine a first order approximation of the metal availability for life.

Current and future missions will rely heavily on our knowledge of the possible mechanisms leading to atmospheric biosignatures in order to determine the presence of Life. Among the most promising possible biosignatures, far from equilibrium atmospheric gases are considered particularly promising \citep{Seager_2014}. 
Compounds such as hydrogen and methane, detected concomitantly with oxidising gases like oxygen or nitrous oxide, suggest the presence of processes sustaining their recycling over time. These gases play a key role in the metabolism of life on Earth. Hydrogen is the most important electron donor in microbial metabolism, and the ability to utilise hydrogen in redox reaction is ubiquitous and probably evolved very early. Similarly methane is a key metabolic by-product of life, and recently the ability to produce methane from a variety of different chemical reactions has been discovered in all known domains of life \citep{Ernst_2022}. 
Molecular oxygen and nitrous oxide are both considered strong biosignatures as on Earth they are exclusively produced through life-controlled reactions (although, molecular oxygen in detectable traces could be due also to abiotic planetary processes \citep{Harman_2018}). 
The enzymes used by life to interact with these gases (either through their production or utilisation) use a diverse set of metals that are uniquely bound to these metabolisms (Table \ref{tab:one}). 
Our hypothesis is that specific biosignatures are likely to be associated with particular metals. For example, hydrogen production and utilisation is possible thanks to NiFe containing enzymes, methane production is nickel and cobalt dependent while oxygen and nitrous oxide utilisation enzymes are dominated by the use of copper containing cofactors. Therefore, the measurement of biometals abundance in exoplanets host-stars will allow a more reliable ranking of the most promising targets in the search for life, but also a thoughtful and more robust evaluation of the potential biosignatures.

A major issue in the search for life by means of astronomical observations is the quantification of the probability (i.e., the Bayesian likelihood) of a positive interpretation of the collected data. Data from a variety of sources need to be considered, given the ambiguous interpretation of any single biosignature \citep{Catling_2018}. The likelihood of the collected data occurring in the presence of life should be weighted by including observed abundance or upper limits on the presence of biometals from the host star.  While the knowledge of the exoplanet parameters (mass, insolation, radius, etc) and its atmosphere and climate strongly informs the evaluation of the posterior Bayesian probability for a positive detection of life, we propose that information about biometals availability should be considered.

The concomitant presence of potential biosignatures and the correlated biometals (according to what we observe on Earth) would strongly support the possibility of a Life instance elsewhere in the Galaxy, and the link between metals and specific metabolism (Table \ref{tab:one}) could help interpret possible atmospheric disequilibria in light of possible metabolisms. Such a concurrent detection would also be very significant as it would support the hypothesis that life elsewhere in the Galaxy would likely have similar basic mechanisms. On the contrary, a very low (i.e., a strong observational upper limit) for biometal content in presence of biosignature would possibly indicate a new path to life. This is an example where astronomy can help testing the universality of biology \citep{Cockell_2018}. The study of exoplanets presents an opportunity to investigate whether Earth's evolutionary path is a singular and chance occurrence, exclusive to our planet. Albeit the strong limitations due the incompleteness of observational data, we think that studying exoplanets would provide a chance to explore whether Earth's evolutionary trajectory is unique and random, limited only to our world.

We suggest that targets with strong deficiency of biometals should be given lower priority in observation follow-up programs searching for biosignatures, as the likelihood to be able to harness or transform energy through these processes might be limited. Clearly, under different environmental conditions other metals might be able to handle the electron flow required to interact with molecules, but overall the ability of metals to mediate redox reactions is a fundamental property of matter, and thus not expected to change even if the instance of life might change significantly in its composition or form. Akin to the requirement for thermodynamic disequilibria, we argue that the necessity to control redox chemistry is a fundamental property of Life itself, thus implying the use of metals in one form or another. 

\section{Conclusions}

It is necessary to move towards a new framework of discussion around the identification of current and future habitability targets. Several authors argue that in order to search for Life in the Solar System and beyond, a conceptually tight proof definition of life and, hence, of habitability, are essential \citep{Vitas_2019}. 
However, we believe that such a definition could be only obtained at the end of our inquiry. In this early stage of our exploration, we need above all a robust and yet pragmatic definition of habitability which can serve us as a guide in selecting and prioritising promising targets. We argued that any notion of habitability must include the requirement of transition elements for life’s metabolism. This is strongly supported by the role played by all biometals in life’s metabolism on Earth. 
%
% OLD: Effectively reducing the space of possibilities for Life requires a well-thought systematic astronomical search. 
%
Effectively reducing the space of possibilities for Life requires a well-designed 
systematic astronomical search, prioritizing
candidates using automated methods analogous to those developed for finding rare phenomena in large surveys \citep[e.g.,][]{Li2021ApJ}.

In the near future, the ESA mission PLATO \citep{Rauer_Plato_2025} will find hundreds of candidate rocky planets in the temperate zone of Sun-like stars. Companion follow-up spectroscopic observations will confirm the exoplanetary interpretation of the transit signals and to determine their mass and bulk density (via Doppler shift technique). 
High-resolution (about a few tens of thousands) spectroscopic follow-ups aimed at measuring the presence of CHNOPS and biometals in the host-stars will be a valuable addition. This also includes planned follow-up observations for measuring radial velocities.

These data will allow the selection of the most promising exoplanet candidates for atmosphere characterization via transmission spectroscopy with the James Webb Space Telescope and the soon-to-come Extremely Large Telescope and the space missions Ariel and Pandora. Ariel and Pandora will be the first space missions entirely dedicated to the study of exoplanetary atmospheres and will likely provide within a decade a census of astrophysical biosignature in nearby exoplanetary systems. 
%In this search we will incur a large number of false positives. 
%
In this search we will incur a large number of false positives.
%, a common difficulty in observational campaigns where potential signals must be carefully distinguished from astrophysical mimics such as variable stars \citep{CalchiNovati2002}.
%
Coupling gas atmospheric disequilibria measurements with information regarding the availability of key metal cofactors (required for their production or utilisation) will provide an additional constraint on the identification of true biosignatures. Future studies regarding the influence of the availability of key metals in controlling the emergence and evolution of specific metabolic pathways on Earth and their role in generating abiotic biosignatures will be required. In this regard, a systematic high-resolution spectroscopic survey of nearby stars and of all the planetary systems host-stars will be a cornerstone in the search for Life in the Galaxy.

%
%  FIGURE 1
%

\begin{figure*}
    \centering
    \includegraphics[width=1\linewidth]{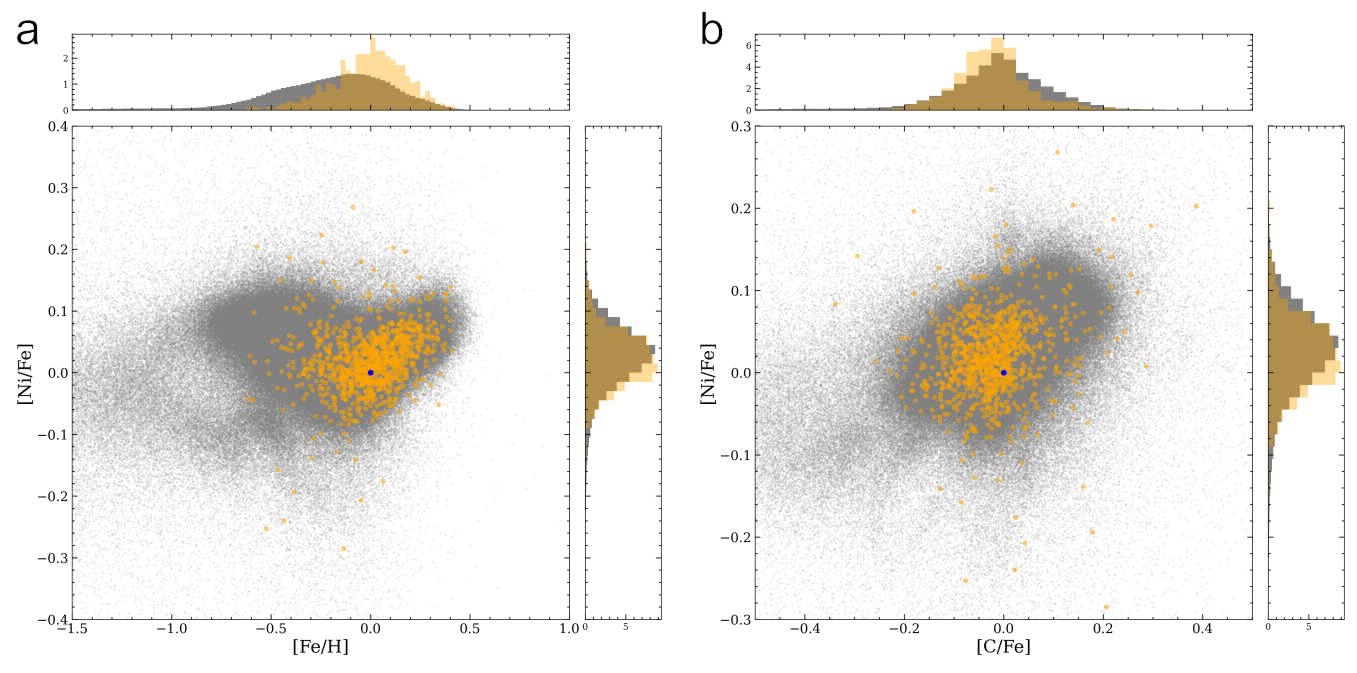}
    \caption{Examples of the distribution of key biometals in the APOGEE star catalogue. The two panels show the abundances of iron, nickel, and carbon for planet-hosting and non-planet hosting stars. Stars with known planets (coloured in orange) span a smaller range in chemical abundances with respect to the general stellar population (grey points). This is likely a consequence of the correlation between stellar metallicity and presence of giant planets. The Sun is coloured blue. Panel a: Relative abundance of nickel as a function of stellar iron-to-hydrogen abundance, sometimes used as a proxy for stellar metallicity. Panel b:, Relative abundance of nickel as a function of the star carbon-to-iron abundance. So far, astronomers focus on metallicity distribution to investigate the connection with formation of planets. Abundance data are essential to inform the selection and interpretation of the most promising targets for search for life.}
    \label{fig:3}
\end{figure*}

\input{table}

\section*{acknowledgments}
This work was supported by 
the University of Napoli Federico II (project: FRA-CosmoHab, CUP E65F22000050001) to GC 
and by 
the European Research Council (ERC) under the European Union’s Horizon 2020 research and innovation programme (grant agreement No. 948972) to DG.

\section*{Data Availability}
No new data was produced for this work.

\appendix
\section{Key definitions at the interface between astronomy and biology}

% This appendix is a contribution to ease the communication effort among the disciplines of astronomy, biology, chemistry and geology.

\textbf{Life vs life.} 
Here % In this perspective 
we refer to Life (with capital L) as a natural phenomenon in its entirety, while we will use life (all lower case) when referring to specific instances of the phenomenon, for example life on Earth.

\textbf{Habitable Zone.} 
Traditionally, the habitable zone is defined as the circumstellar region where a planet with an Earth-like atmosphere will have surface liquid water. This term can be misleading, as the presence of liquid water and the potential to host life only loosely overlap. The so-called “habitable zone” is just a first order approximation of the effective habitability of a planet. More nuanced definitions have been proposed for the habitable zone \citep{Tasker_2017}, and the region where surface liquid water can be present is more appropriately termed “temperate zone”, or as suggested by other authors,  the “liquid water belt.” In order to build a robust measure of exoplanets' potential to support life, a more comprehensive definition of habitability is needed.

\textbf{Trace elements.} 
Trace elements are operationally defined as the elements present in small quantities (a "trace amount"). This definition effectively changes the list of trace elements depending on the frame of reference. For example, in biology these are typically defined as all the elements necessary for the growth of organisms excluding the main building blocks elements CHNOPS, while in geology they are defined with respect to rock forming elements.

\textbf{Metals.} 
In astronomy, metals are broadly defined as all elements with atomic numbers larger than 2. In chemistry, metals are elements that readily form cations in solution, are good heat and electricity conductors and form metallic bonds. Metals are further divided in alkali metals, alkaline metals, lanthanides, actinides, transition metals and metalloids.

\textbf{Metallicity.}
The stellar metallicity is commonly represented by the quantities $Z$ and [M/H].
$Z$ is the total mass fraction of elements heavier than helium in a star.
[M/H] is the logarithm of the ratio of the number density of all such elements to hydrogen, relative to the Sun.

\textbf{Biometals.} 
Biometals are defined as a subgroup of metals (as defined in chemistry) that have specific biological roles as cofactors with regard to protein structure and function \citep{Giovannelli_2023, Hay_Mele_2023}. 

\textbf{Cofactors.} 
Cofactors are organic or inorganic molecules that interact with the enzyme to activate or speed up (i.e., catalyse), a chemical reaction.

\textbf{Enzymes.} A protein, or protein complex, capable of catalysing a set of specific chemical reactions.

\textbf{Oxidoreductases.} A specific group of enzymes involved in oxido-reduction reactions.

\textbf{Biomass.} The total mass of organisms in a given area or volume, often expressed in terms of amount of carbon. Biomass is created through metabolic processes that are fueled either by the breakdown of organic carbon compounds (heterotrophy) or the synthesis of organic compounds from inorganic precursors (autotrophy).

\textbf{Redox reactions. }Redox reactions (from REDuction-OXidation reactions) are a chemical process in which the oxidation state of one or more atoms involved changes due to the exchange of electron between a donor (called reductant, that gets oxidised) and an acceptor (called an oxidant, that gets reduced). Redox chemistry is important in biology, material science, energy generation and a large number of natural processes. 

\section{Redox chemistry and energy in living systems}

Life uses a large number of chemical elements to build up its biomass. The main macromolecules composing life biomass are lipids, proteins, carbohydrates and nucleic acids. All these macromolecules are composed for the vast majority by a small number of elements known as life’s building blocks. These elements, namely carbon, hydrogen, nitrogen, oxygen, phosphorus and sulphur (often referred to as the CHNOPS elements from their chemical formula) on average constitute up to 98.9\% of an organism's dry weight. The availability of these elements is considered essential for the emergence of Life, and their availability across time and space can limit the amount of biomass produced. In addition to CHNOPS elements, living organisms need a large number of other elements that appear in minor abundance (trace elements) within molecules and organic and inorganic cofactors. This is due to the crucial role that trace elements have in chemical reactions that Life uses to obtain energy and build up biomass \citep{Giovannelli_2023}.

Life on our planet is able to harvest energy from a diverse array of thermodynamic disequilibria, generally in the form of redox chemistry. This is generally accomplished by diverting ions (mainly as protons) and electrons through different paths in the cell, ultimately allowing to build up electrochemical potential used to store energy and carry out work. The amount of different redox couples that life is able to utilise has been estimated to be in the range of several thousands of reactions, to which we need to add the ability of life to extract electrons using light from a broad wavelength range (i.e., phototrophy). Each of these reactions has a specific midpoint redox potential that life on Earth is able to access thanks to specialised proteins called oxidoreductases, that constitute life’s engines. These proteins, generally, have one or more metal cofactors that directly act as catalytic centres for exchanging electrons. The exact range of redox potential available to each oxidoreductase is controlled in a variety of different ways, starting with the use of a diverse array of transition metals that include Fe, Cu, Mo, V, W, Co, Ni, Mg, Mn among others \citep{Giovannelli_2023} (Figure 1). Recently, the role of trace elements in controlling metabolisms has come into focus (see \cite{Giovannelli_2023} for a review), and diverse initiatives are honing on the role of trace elements in controlling life emergence and evolution.

\bibliographystyle{mnras}

\bibliography{references}

\end{document}

%% file: table.tex
\begin{table*}
\centering
\caption{List of potential biosignature gases, their notes, Earth metabolism pathways, and metals involved in the pathway}
\begin{tabular}{>{\centering\arraybackslash}m{2.5cm}|>{\arraybackslash}m{5.5cm}|>{\arraybackslash}m{5cm}|>{\arraybackslash}m{3cm}}
\hline
\thead{Biosignature gas} & \thead{Notes} & \thead{Earth Metabolism} & \thead{Metals involved \\ in the pathway} \\ 
\hline
CH\(_4\) & 

Methane, an important greenhouse gas and potentially a biosignature if present with strong oxidants (e.g., oxygen)&
Methanogenesis (production) 

& Ni, Fe, Co, W, Mo, Cu 

\\ 
\cline{3-4} 
& & Methanotrophy (consumption) & Fe, Cu, Mo, Ni, W \\ 
\hline
O\(_2\) & Oxygen, a strong oxidant produced on Earth through oxygenic photosynthesis. O\(_3\) can be used as an indicator of oxygen presence & 
Oxygenic photosynthesis (production)

& Mn, Fe, Cu, Mg 

\\ 
\cline{3-4}
& &

Aerobic respiration (consumption)

&

Fe, Cu, Mn, Zn 

\\ 
\hline
N\(_2\)O 

&
Nitrous oxide, possible biosignature if present together with reductants (CH\(_4\), H\(_2\)) in the atmosphere & Denitrification, nitrous oxide reduction (consumption) 
\vspace{1mm} & 
Fe, Cu, Mn, Ni \\ 
\hline
NH\(_3\) & 
Ammonia, possible biosignature in the atmosphere since detectable levels have to be sustained by a potentially biological surface production & Denitrification to ammonia, nitrogen fixation (production) 
\vspace{1mm} & 
Mo, Fe, V 

\\ 
\cline{3-4}
& & Ammonia oxidation (consumption) & 
Cu, Fe, Mo, Ni \\ 
\hline
(CH\(_3\))\(_2\)S, CH\(_3\)Cl, CH\(_3\) SH & Methyl containing gases, produced on Earth through biological pathways might be established as alternative biosignatures & 

Carbon fixation (production), secondary metabolite production, halogenation

& 
Fe, Mo, V,
C, Cu

\\ 
\hline

H\(_2\)O 
\vspace{2mm}
& 
Water, used as an indicator of habitable conditions

& 
Used as electron donor in oxygenic photosynthesis, produced as a byproduct of heterotrophic aerobic (oxygen consuming) respiration 
& 
Mn, Fe, Cu 
\vspace{4mm}

\\ 
\hline
\end{tabular}
\label{tab:one}
\end{table*}